\documentclass[prl,notitlepage,twocolumn]{revtex4}

\usepackage{amsmath}
\usepackage{amssymb}
\usepackage{bm}
\usepackage{graphicx}
\usepackage{times}
\usepackage{xcolor}
\usepackage{float}
\usepackage{gensymb}

\usepackage{pdfcomment}
\pdfcommentsetup{draft} % set to 'final' to prevent all comments from showing up
\pdfcommentsetup{color={1.0 1.0 0.0}, open=true}

\usepackage{color}
\usepackage{ulem}

\begin{document}

\title{Asymmetric metasurfaces with high-$Q$ resonances governed by bound states in the continuum}

\author{Kirill Koshelev$^{1,2}$}
\author{Sergey Lepeshov$^2$}
\author{Mingkai Liu$^1$}
\author{Andrey Bogdanov$^2$}
\author{Yuri Kivshar$^{1,2}$}
\affiliation{$^1$Nonlinear Physics Centre, Australian National University, Canberra ACT 2601, Australia}
\affiliation{$^2$ITMO University, St. Petersburg 197101, Russia}

\begin{abstract}
We reveal that metasurfaces created by seemingly different lattices of (dielectric or metallic) meta-atoms with broken in-plane symmetry can support sharp high-$Q$ resonances that originate from the physics of bound states in the continuum.  We prove rigorously a direct link between the bound states in the continuum and the Fano resonances, and develop a general theory of such metasurfaces,
suggesting the way for smart engineering of resonances for many applications in nanophotonics and meta-optics.
\end{abstract}

\maketitle

Metasurfaces have attracted a lot of attention in the recent years due to novel ways for wavefront control, advanced light focusing, and ultra-thin optical elements~\cite{capasso}. Recently, metasurfaces based on high-index resonant dielectric materials~\cite{kuznetsov2016optically} have emerged as essential building blocks for various functional meta-optics devices~\cite{kruk2017functional} due to their low intrinsic loss, with unique capabilities for controlling the propagation and localization of light. A key concept underlying the specific functionalities of many metasurfaces is the use of constituent elements with spatially varying optical properties and optical response characterized by high quality factors ($Q$ factors) of the resonances.

\begin{figure}[t]
   \centering
   \includegraphics[width=0.98\columnwidth]{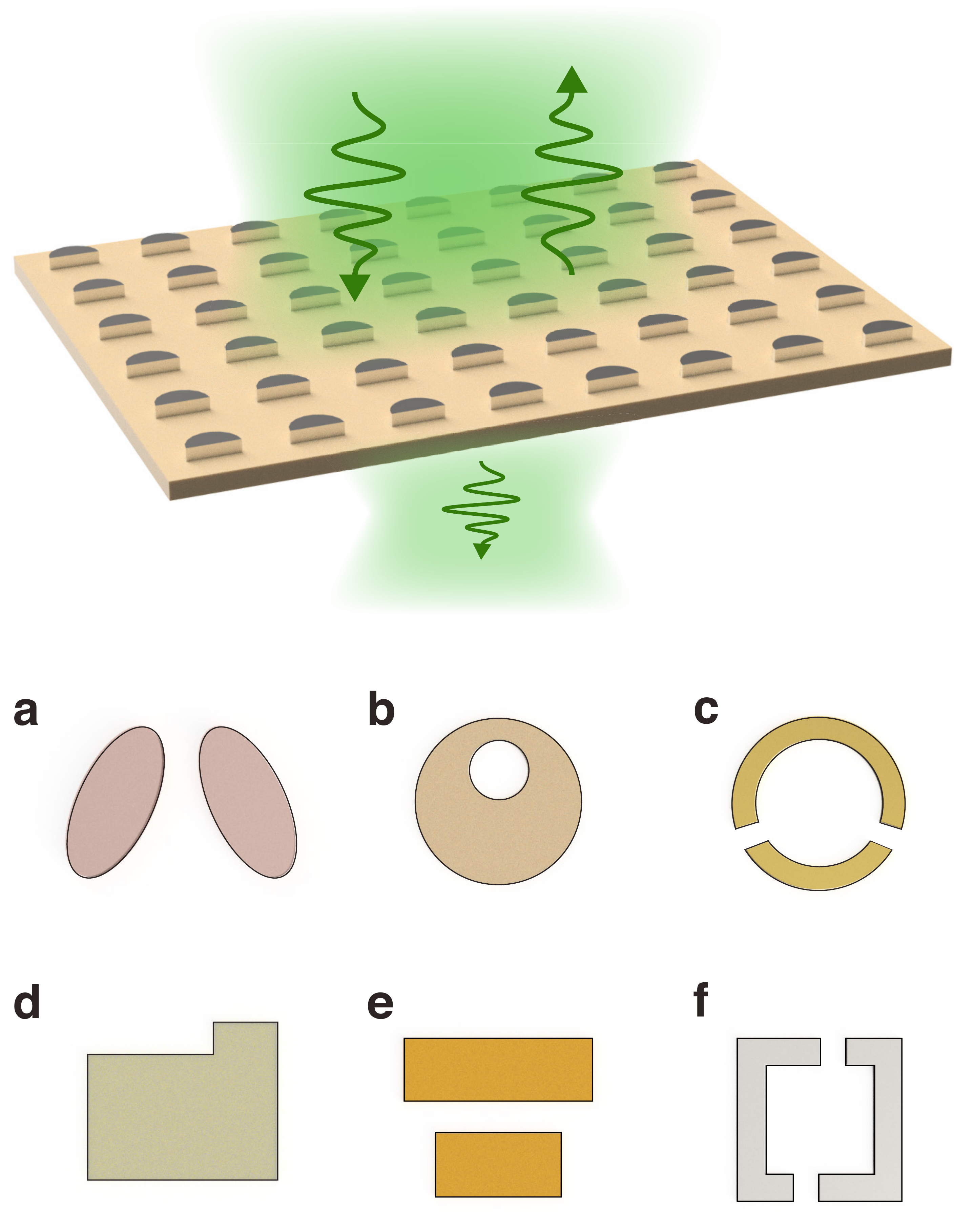} % requires the graphicx package
   \caption{Top: Schematic of light scattering by a metasurface. Bottom: Designs of unit cells of metasurfaces with a broken in-plane inversion symmetry of constituting meta-atoms supporting sharp resonances, analysed in  Refs.~\cite{science,Mingkai,tuz2018high,fedotov2007sharp,forouzmand2017all,campione2016broken, vabishchevich2018enhanced,lim2018universal,khardikov2010trapping, singh2011observing}.}
   \label{fig:figure_1}
\end{figure}

Many interesting phenomena have been shown for metasurfaces composed of arrays of meta-atoms with broken in-plane inversion symmetry (see Fig.~\ref{fig:figure_1}), which all demonstrate the excitation of high-$Q$ resonances for the normal incidence of light. The examples are the demonstration of imaging-based molecular barcoding with pixelated dielectric metasurfaces~\cite{science} and manifestation of polarization-induced chirality in metamaterials~\cite{Mingkai}, which both are composed of asymmetric pairs of tilted bars [see Fig.~\ref{fig:figure_1}(a)], observation of trapped modes in arrays of dielectric nanodisks with asymmetric holes~\cite{tuz2018high} [see Fig.~\ref{fig:figure_1}(b)], sharp trapped-mode resonances in plasmonic and dielectric split-ring structures~\cite{fedotov2007sharp,forouzmand2017all} [see, e.g.,  Fig.~\ref{fig:figure_1}(c)], broken-symmetry Fano metasurfaces for enhanced nonlinear effects~\cite{campione2016broken, vabishchevich2018enhanced} [see Fig.~\ref{fig:figure_1}(d)], tunable high-$Q$ Fano resonances in plasmonic metasurfaces~\cite{lim2018universal} [see Fig.~\ref{fig:figure_1}(e)], trapped light and metamaterial-induced transparency in arrays of square split-ring resonators~\cite{khardikov2010trapping, singh2011observing} presented in Fig.~\ref{fig:figure_1}(f). Here, we demonstrate that all such seemingly different structures can be unified by a general concept of bound states in the continuum, and we prove rigorously their link to the Fano resonances.

Bound states in the continuum (BICs) originated from quantum mechanics as a curiosity~\cite{von1929uber}, but later they were rediscovered as an important physical concept of destructive interference~\cite{Friedrich1985} being then extended to other fields of wave physics, including acoustics~\cite{parker1966resonance} and optics~\cite{Marinica2008,Bulgakov2008}. A true BIC is {\it a mathematical object} with an infinite $Q$ factor and vanishing resonance width, it can exist only in ideal lossless infinite structures or for extreme values of parameters~\cite{Ndangali2010, Hsu2013, Monticone2014}. In practice, BIC can be realized as a quasi-BIC in the form of {\it a supercavity mode}~\cite{rybin2017optical} when both $Q$ factor and resonance width become finite at the BIC conditions due to absorption, size effects~\cite{Balezin}, and other perturbations~\cite{Sadrieva2017}. Nevertheless, the BIC-inspired mechanism of light localization makes possible to realize high-$Q$ {\it quasi-BIC} in optical cavities and photonic crystal slabs~\cite{Bulgakov2008,Hsu2013,rybin2017optical}, coupled optical waveguides~\cite{Plotnik2011,molina2012surface,corrielli2013observation}, and even isolated subwavelength dielectric particles~\cite{rybin2017high}.

\begin{figure*}[t]
   \centering
   \includegraphics[width=1.9\columnwidth]{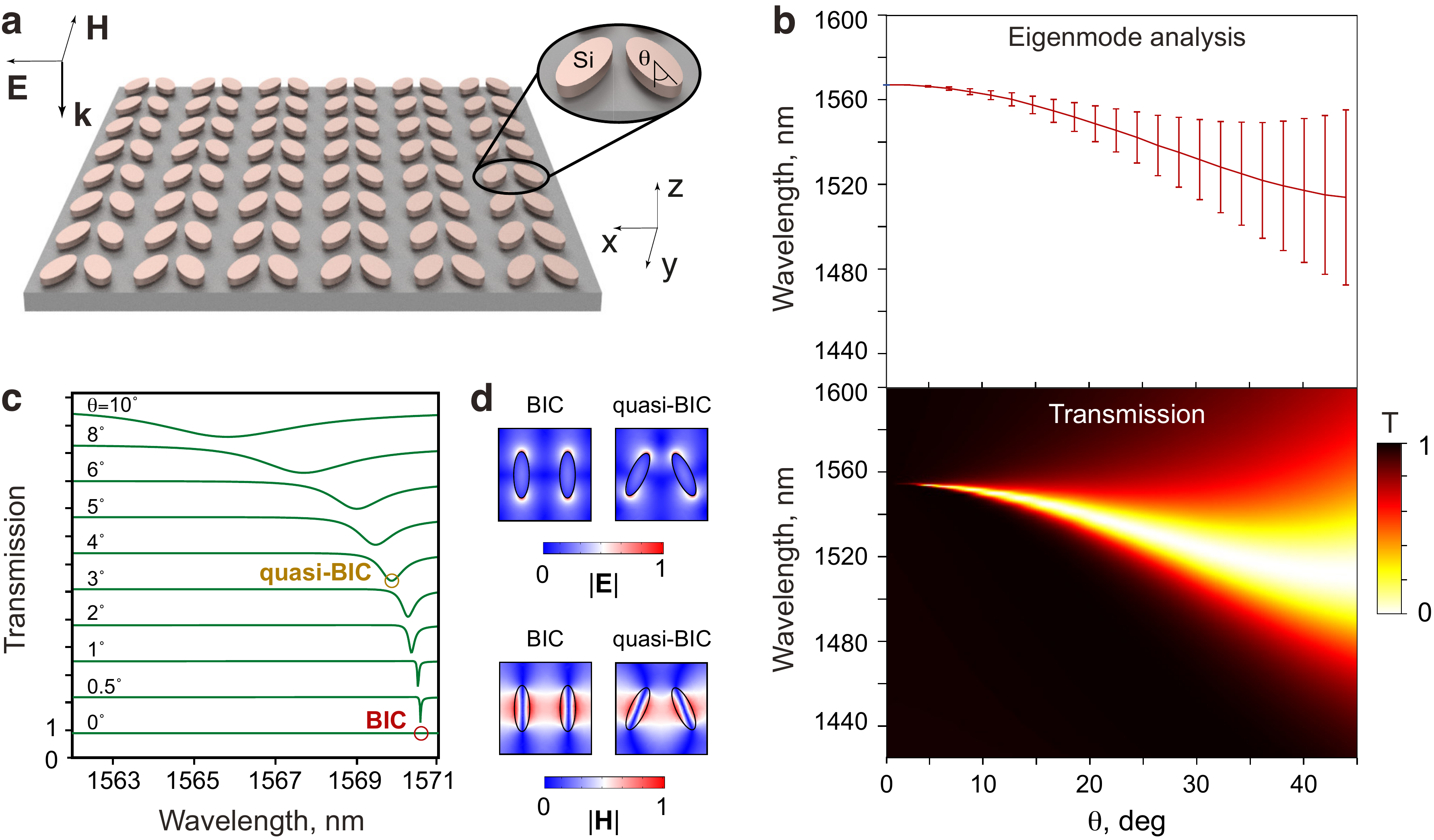} % requires the graphicx package
   \caption{(a) A square lattice of tilted silicon-bar pairs with a design of the unit cell. Parameters: the period is 1320 nm, bar semi-axes are 330 nm and 110 nm, height is 200 nm, distance between bars is 660 nm. (b) Eigenmode spectra and transmission spectra with respect to pump wavelength and angle $\theta$. Error bars show the magnitude of the mode inverse radiation lifetime.  (c) Evolution of the transmission spectra vs. angle  $\theta$. Spectra are relatively shifted by 1.5 units. (d) Distribution of the electric and magnetic fields for BICs and quasi-BICs. }
   \label{fig:figure_2}
\end{figure*}

In this Letter, we reveal that sharp spectral resonances recently reported and even observed for many types of seemingly different plasmonic and dielectric metasurfaces originate in fact from the powerful concept of BIC, and this finding allows us to predict and engineer high-$Q$ resonances in nanophotonics. We demonstrate that true BICs transform into quasi-BICs  when the in-plane inversion symmetry of a unit cell becomes broken, and we derive the universal formula for the $Q$ factor as a function of the asymmetry parameter. We develop a novel analytical approach to describe light scattering by arrays of meta-atoms based on the explicit expansion of the Green's function of an open system into eigenmode contributions, and demonstrate rigorously that reflection and transmission coefficients are linked to the conventional Fano formula. We prove that the Fano parameter becomes ill-defined at the BIC condition which corresponds to the collapse of the Fano resonance~\cite{Fonda1963,kim1999cs}. Our findings pave the way towards a novel approach to the engineering of resonances in meta-optics and nanophotonics.

To gain a deeper insight into the physics of BICs in metasurfaces with in-plane asymmetry, we focus on the design recently suggested for biosensing~\cite{science}, namely, a square array of tilted silicon-bar pairs shown in Fig.~\ref{fig:figure_2}(a). For calculations, we consider a homogenous background with permittivity $1$. We calculate both the eigenmode and transmission spectra as shown in Fig.~\ref{fig:figure_2}(b). An asymmetry parameter is introduced through the angle $\theta$ between the $y$-axis and the long axis of the bar. The ideal (lossless and infinite) structure supports a true optical BIC at $\theta = 0^{\degree}$~\cite{zhen2014topological}. Such an ideal BIC is unstable against perturbations that break the in-plane symmetry $(x,y)\rightarrow (-x,-y)$, and it transforms into a quasi-BIC  with a finite $Q$ factor.

The eigenmode and transmission spectra are shown in Fig.~\ref{fig:figure_2}(b) as functions of $\theta$ , where $T=|t|^2$ is the transmittance and $t$ is the amplitude of the transmitted wave. We observe that the BIC with infinite $Q$ factor at $\theta=0^{\degree}$ transforms into a high-$Q$ quasi-BIC, whose radiation loss increases with $\theta$. The detailed transmission spectra shown in Fig.~\ref{fig:figure_2}(c) exhibit a narrow dip that vanishes when the pair becomes symmetric, which confirms the results of the eigenmode analysis. Figure~\ref{fig:figure_2}(d) demonstrates similarity of the distributions of electric and magnetic fields in BICs and quasi-BICs within a unit cell. Analysis shows that the BIC carries a topological charge of 1 (see Supplemental Material).

We analyze the transmission spectra of such metasurfaces and prove rigorously that it can be described by the classical Fano formula, and the observed peak position and linewidth correspond exactly to the real and imaginary parts of the frequency of the eigenmode. While the analytical solution of Maxwell's equations does not exist, the description of the transmission $T$ with the Fano formula is still widely used by introducing the Fano parameter phenomenologically~\cite{fano}. Here, we demonstrate the {\it explicit} correspondence between the Fano lineshape of the transmission spectra and properties of the eigenmode spectra. It is worth mentioning that previously the Fano formula for BICs in periodic photonic structures was dicussed or obtained only for special assumptions~\cite{Marinica2008, Hsu2013, fano2016}.

\begin{figure}[t]
   \centering
   \includegraphics[width=0.95\columnwidth]{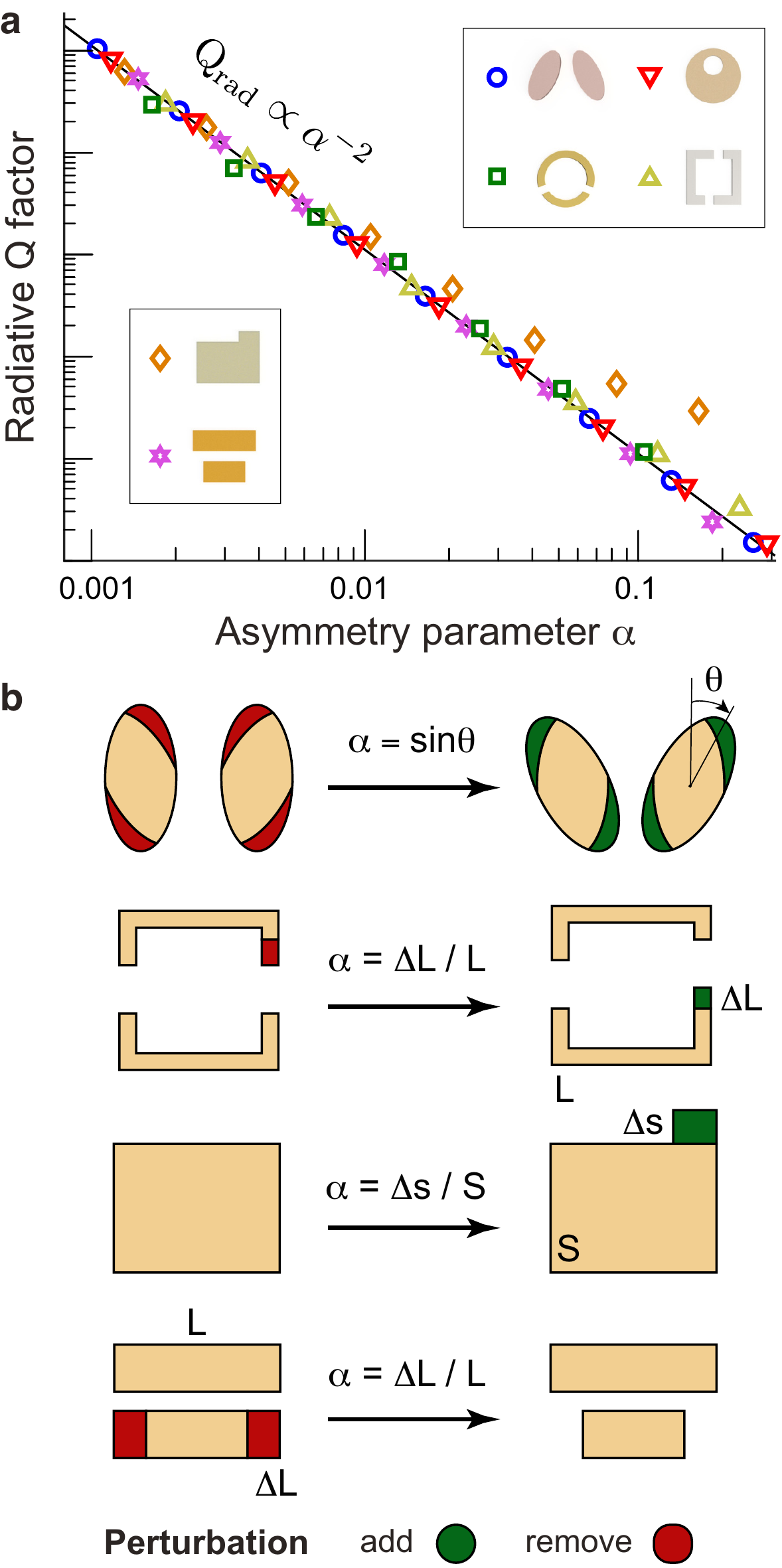} % requires the graphicx package
   \caption{ Effect of in-plane asymmetry on the radiative $Q$ factor of quasi-BICs. (a) Dependence of the $Q$ factor on the asymmetry parameter $\alpha$ for all designs of symmetry-broken meta-atoms shown in Figs.~\ref{fig:figure_1}(a-f), in log-log scale. (b) Definitions of the asymmetry parameter $\alpha$ for some of the structures.}
   \label{fig:figure_3}
\end{figure}

To derive an analytical expression for light transmission, we expand the transmitted field amplitude into a sum of independent terms where each term corresponds to an eigenmode of the photonic structure. This becomes possible by applying the recently developed procedure allowing for rigorous characterization of the Green's function and, therefore, the eigenmode spectra of open optical resonators~\cite{muljarov2017}. The eigenmodes of a metasurface are treated as self-standing resonator excitations with a complex spectrum describing both the resonant frequencies  $\omega_0$ and inverse lifetimes $\gamma$. Straightforward but rather cumbersome calculations (see Supplemental Material) reveal that the frequency dependence of the transmission $T$ is described rigorously by the Fano formula, and the Fano parameters are expressed explicitly through the material and geometrical parameters of the metasurface and dimensionless frequency $\Omega = 2(\omega-\omega_0)/\gamma$,
\begin{equation}
T(\omega) = \frac{T_0}{1+q^2}\frac{(q+\Omega)^2}{1+\Omega^2} + T_{\rm bg}(\omega).
\label{eq:1}
\end{equation}
Here $q$ is the Fano asymmetry parameter, $T_0$ and $T_{\rm bg}$ describe the smooth background contribution of non-resonant modes to the resonant peak amplitude and the offset, respectively (see the explicit expressions in Supplemental Material~\cite{supp}). Remarkably, the exact formula shows that the parameter $q$ in Eq.~(\ref{eq:1}) for BICs supported by metasurfaces with unbroken in-plane symmetry becomes ill-defined, which corresponds to a collapse of the Fano resonance when any features in the transmission spectra disappear, and the resonant mode is transformed into a $"$dark mode$"$.

Next, we describe analytically the behavior of the radiative $Q$ factor of the quasi-BIC as a function of $\theta$ shown in Figs.~\ref{fig:figure_2}(b-c). We consider the radiation losses as a perturbation which is a natural approximation valid when $\theta$  remains relatively small. Then, the inverse radiation lifetime $\gamma_{\rm rad}$ can be calculated as a sum of radiation losses into all open radiation channels. We focus on the quasi-BICs with the frequencies below the diffraction limit for which only open radiation channels represent the zeroth-order diffraction. Then $\gamma_{\rm rad}$ of the quasi-BIC takes the form (see Supplemental Material),
\begin{subequations}
\begin{align}
&\frac{\gamma_{\rm rad}}{c} = |D_x|^2+|D_y|^2,\\
D_{x,y} = -\frac{k_0}{\sqrt{2S_0}} & \left[p_{x,y}\mp \frac{1}{c} {m_{y,x}}+\frac{ik_0}{6}Q_{xz,yz} \right].
\end{align}
\label{eq:Fermi}
\end{subequations}
Here $k_0=\omega_0/c$, $S_0$ is the unit cell area,  $\mathbf{p}$, $\mathbf{m}$ and $\mathbf{\hat Q}$ are the electric dipole, magnetic dipole and electric quadrupole normalized moments of a unit cell, which depend on $\theta$.
\begin{figure}[t]
   \centering
   \includegraphics[width=0.97\columnwidth]{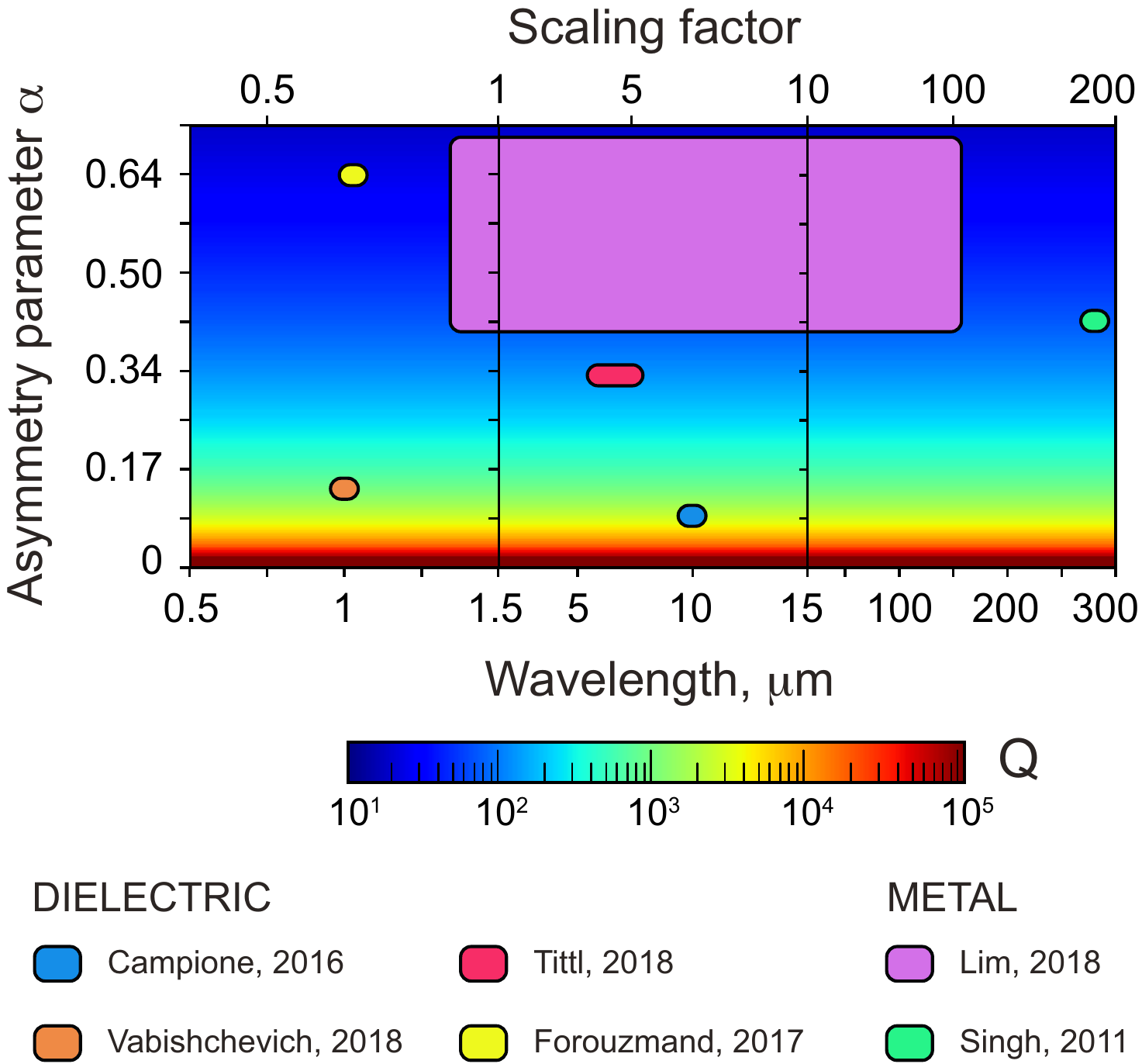} % requires the graphicx package
   \caption{Map of operating wavelengths and quality factors for silicon metasurfaces with tilted-bar pairs by varying the orientation of the bars ($\alpha = \sin{\theta}$) and by scaling the metasurface. The radiative part of the total Q factor is evaluated using Eq.~(\ref{eq:law}), the non-radiative part is taken equal $10^5$. All calculations are confirmed by direct numerical simulations with realistic dispersion of silicon (see Supplemental Material). Geometric scaling factor is shown in the upper horizontal axis. Colored rectangles correspond to the structures considered in the previous studies, see Fig.~\ref{fig:figure_1}.}
   \label{fig:figure_4}
\end{figure}
The amplitudes $D_{x,y}$ govern the overlap coefficients between the mode profile inside the photonic structure and the profile of vertically propagating free-space modes of two orthogonal polarizations, respectively.  Thus, Eqs.~(\ref{eq:Fermi}a-b) show that for a true BIC mode symmetry mismatch leads to zero $D_{x,y}$ and vanishing radiation losses~\cite{lee2012observation}. In other words, the electric field components $E_x, E_y$ of a BIC are odd with respect to the inversion of coordinates $(x,y)\rightarrow (-x,-y)$ so $\gamma_{\rm rad}=0$. For quasi-BICs, we perform straightforward transformations of Eqs.~(\ref{eq:Fermi}a-b) (see Supplemental Material) to show that the radiative quality factor $Q_{\rm rad}=\omega_0/\gamma_{\rm rad}$ depends on $\theta$ as
\begin{equation}
Q_{\rm rad}(\theta) =  Q_0\left[\alpha(\theta)\right]^{-2},
\label{eq:law}
\end{equation}
where $\alpha = \sin{\theta}$ and $Q_0$ is a constant independent on $\theta$. In general, Eq.~(\ref{eq:law}) remains valid for metasurfaces placed on a substrate as long as the quasi-BIC frequency is below the diffraction limit of the substrate~\cite{Sadrieva2017}.

Next, we demonstrate that the quadratic dependence defined by Eq.~(\ref{eq:law}) represents a universal behavior of the $Q$ factor of a quasi-BIC mode as a function of the asymmetry parameter for all dielectric and plasmonic metasurfaces with broken-symmetry meta-atoms. We introduce the generalised asymmetry parameter $\alpha$, which has distinct definitions for different structures but takes values between 0 and 1. We derive Eq.~(\ref{eq:law}) using the second-order perturbation theory for open electromagnetic systems and confirm the result by independent calculations of the eigenmode and reflectance spectra for all designs presented in Figs.~\ref{fig:figure_1}(b-f) (see Supplemental Material). Since plasmonic metasurfaces possess significant absorption, we extract the bare radiative $Q$ factor by evaluating the inverse radiative lifetime $\gamma_{\rm rad}$ being a difference between the total inverse lifetime $\gamma$ and non-radiative damping rate evaluated at $\alpha=0$.

Figure~\ref{fig:figure_3}(a) shows a direct comparison of the values of the radiative $Q$ factor of quasi-BICs as functions of the asymmetry parameter $\alpha$ for dielectric and plasmonic metasurfaces with various broken-symmetry meta-atoms in the unit cell. All curves are shifted relatively in the vertical direction to originate from the same point. As can be noticed from Fig.~\ref{fig:figure_3}(a), for small values of $\alpha$ the behavior of  $Q_{\rm rad}$  for all metasurfaces is clearly inverse quadratic. Importantly, for most of the structures the law $\alpha^{-2}$   is valid beyond applicability of the perturbation theory. Figure~\ref{fig:figure_3}(b) introduces the definition of the asymmetry parameter $\alpha$ for different metasurface designs. 
%These include two varieties of the symmetry-broken meta-atoms defined by the type of the transformation that breaks the structure symmetry. In the first type, one does not add any new material but only changes a shape, whereas for another type the symmetry is broken by adding or removing some material. 

The quadratic scalability of the $Q$ factor of quasi-BICs combined with linear scalability of Maxwell's equations suggests a straightforward way of smart engineering of photonic structures with the properties on demand. As an example, we focus on a design of tilted silicon-bar pairs and suggest a very simple algorithm for a design of metasurfaces with a wide range of $Q$ factors and operating wavelengths ranging from from visible to THz.  First, since dispersion of the refractive index for silicon is relatively weak, we can tune the operating wavelength from 0.5 $\mu$m to 300 $\mu$m by a linear geometric scaling of the structure in all dimensions. Second, we can control the mode radiative $Q$ factor in a wide range of parameters by changing the asymmetry parameter $\alpha = \sin{\theta}$ according to Eq.~(\ref{eq:law}). The total Q factor of a quasi-BIC is limited by absorption, which can be estimated as ${\rm Re}(\varepsilon)/{\rm Im}(\varepsilon)$. For silicon in the range from near-IR to THz, it is more than $10^5$.

Using this approach, we calculate the dependence of total $Q$ factor of the quasi-BIC vs. operating wavelength and asymmetry parameter for a square lattice of tilted silicon-bar pairs, and summarize the results in Fig.~\ref{fig:figure_4}. For all calculations done for this map, we make only one numerical simulation with fixed material parameters to obtain the value of $Q_0$ required for use of Eq.~(\ref{eq:law}), then exploit the advantages of the geometrical scaling combined with rotation of bars. {\it We observe that the values of available $Q$ factors can be achieved in a broad range from $10$ up to $10^5$} for each wavelength domain, by using the same material and design. The colored rectangles show the range of $Q$ factors and operating wavelengths reported in the previous studies of metasurfaces with symmetry-broken meta-atoms~\cite{science,forouzmand2017all,campione2016broken, vabishchevich2018enhanced, lim2018universal, singh2011observing}. We verify the applicability of the proposed analytical approach by three-dimensional electromagnetic simulations with the silicon dispersion by using the finite-element method in COMSOL, and the results are in Supplemental Material.  Both approaches agree well, thus justifying the validity of our analytical scaling method.

We believe our approach based on the BIC concept can describe many other cases of symmetry-broken metasurfaces and also photonic crystral slabs, studied earlier with different applications in mind~\cite{thomas,shvets,costas2,tian, shanhui}. Also, our approach can be helpful to get a deeper physical insight into many other problems in optics, including dark bound states in dielectric inclusions coupled to the external waves by small non-resonant metallic antennas~\cite{costas} and electromagnetically induced transparency~\cite{eit}. We argue that almost any problem involving the so-called $"$dark states$"$ can find its rigorous formulation with the powerful theory of BIC resonances.

In conclusion, we have demonstrated that  high-$Q$ resonances recently observed in metasurfaces composed of dissimilar meta-atoms with  broken in-plane inversion symmetry are associated with the concept of bound states in the continuum. We have proven rigorously a direct link between peculiarities of the transmission spectra, Fano resonances, and the existence of high-$Q$ quasi-BIC resonances. We have explained analytically the variation of the $Q$ factor with a change of the unit-cell asymmetry that paves the way towards smart engineering of sharp resonances in meta-optics for nanolasers, light-emitting metasurfaces, optical sensors, and ultrafast active metadevices.

\begin{acknowledgments}
The authors acknowledge a financial support from the Australian Research Council, the AvH Foundation and the Russian Science Foundation (17-12-01581), and useful discussions with H. Altug, H. Atwater, F. Capasso, A. Krasnok, S. Kruk, Th. Pertsch, M. Rybin, R. Singh,  V. Tuz and N. Zheludev.
\end{acknowledgments}


\begin{thebibliography}{99}

\bibitem{capasso}
N. Yu and F. Capasso, Flat optics with designer metasurfaces, Nat. Mater. {\bf 13}, 139 (2014).

\bibitem{kuznetsov2016optically}
A. I. Kuznetsov, A. E. Miroshnichenko, M. L. Brongersma, Y. S. Kivshar, and B. Luk’yanchuk, Optically resonant dielectric nanostructures, Science {\bf 354}, aag2472 (2016).

\bibitem{kruk2017functional}
S. Kruk and Y. Kivshar, Functional meta-optics and nanophotonics governed by Mie resonances, ACS Photonics {\bf 4}, 2638 (2017).

\bibitem{science}
A. Tittl, A. Leitis, M. Liu, F. Yesilkoy, D.Y. Choi, D.N. Neshev, Y.S. Kivshar, and H. Altug, Imaging-based molecular barcoding with pixelated dielectric metasurfaces, Science {\bf 360}, 1105 (2018).

\bibitem{Mingkai}
M. Liu, D. A. Powell, R. Guo, I. V. Shadrivov and Y. S. Kivshar, Polarization-induced chirality in metamaterials via optomechanical interaction, Adv. Opt. Mater. {\bf 5}, 1600760 (2017).

\bibitem{tuz2018high}
V. R. Tuz, V. V. Khardikov, A. S. Kupriianov, K.L. Domina, S. Xu, H. Wang, and H.-B. Sun, High-quality trapped modes in all-dielectric metamaterials, Opt. Express {\bf 26}, 2905 (2018).

\bibitem{fedotov2007sharp}
V. A. Fedotov, M. Rose, S. L. Prosvirnin, N. Papasimakis, and N. I. Zheludev, Sharp trapped-mode resonances in planar metamaterials with a broken structural symmetry, Phys. Rev. Lett. {\bf 99}, 147401 (2007).

\bibitem{forouzmand2017all}
A. Forouzmand and H. Mosallaei, All-dielectric $C$-shaped nanoantennas for light manipulation: Tailoring both magnetic and electric resonances to the desire, Adv. Opt. Mater. {\bf 5}, 1700147 (2017).

\bibitem{campione2016broken}
S. Campione, S. Liu, L. I. Basilio, L. K. Warne, W. L. Langston, T. S. Luk, J. R. Wendt, J. L. Reno, G. A. Keeler, I. Brener, and M. B. Sinclair, Broken symmetry dielectric resonators for high quality factor Fano metasurfaces, ACS Photonics {\bf 3}, 2362 (2016).

\bibitem{vabishchevich2018enhanced}
P. P. Vabishchevich, S. Liu, M. B. Sinclair, G. A. Keeler, G. M. Peake, and I. Brener, Enhanced second-harmonic generation using broken symmetry III-V semiconductor Fano metasurfaces, ACS Photonics {\bf 5}, 1685 (2018).

\bibitem{lim2018universal}
W.X. Lim and R. Singh, Universal behaviour of high-Q Fano resonances in metamaterials: Terahertz to near-infrared regime, Nano Convergence {\bf 5}, 5 (2018).

\bibitem{khardikov2010trapping}
V.V. Khardikov, E.O. Iarko, and S.L. Prosvirnin, Trapping of light by metal arrays, J. Opt. {\bf 12}, 045102 (2010).

\bibitem{singh2011observing}
R. Singh, I.A. Al-Naib, Y. Yang, D. Roy Chowdhury, W. Cao, C. Rockstuhl, T. Ozaki, R. Morandotti, and W. Zhang, Observing metamaterial induced transparency in individual Fano resonators with broken symmetry, Appl. Phys. Lett. {\bf 99}, 201107 (2011).

\bibitem{von1929uber}
J. Von Neuman and E. Wigner, \"{U}ber merkw\"{u}rdige diskrete Eigenwerte, Phys. Z. {\bf 30}, 467 (1929).

\bibitem{Friedrich1985}
H. Friedrich and D. Wintgen, Interfering resonances and bound states in the continuum, Phys. Rev. A {\bf 32}, 3231 (1985).

\bibitem{parker1966resonance}
R. Parker, Resonance effects in wake shedding from parallel plates: some experimental observations, J. Sound Vibr. {\bf 4}, 62 (1966).

\bibitem{Marinica2008}
D.C. Marinica, A.G. Borisov, and S.V. Shabanov, Bound states in the continuum in photonics, Phys. Rev. Lett. {\bf 100}, 183902 (2008).

\bibitem{Bulgakov2008}
E.N. Bulgakov and A.F. Sadreev, Bound states in the continuum in photonic waveguides inspired by defects, Phys. Rev. B {\bf 78}, 075105 (2008).

\bibitem{Ndangali2010}
R.F. Ndangali and S.V. Shabanov, Electromagnetic bound states in the radiation continuum for periodic double arrays of subwavelength dielectric cylinders, J. Math. Phys. {\bf 51}, 102901 (2010).

\bibitem{Hsu2013}
C.W. Hsu, B. Zhen, J. Lee, S.-L. Chua, S.G. Johnson, J.D. Joannopoulos, and M. Solja\^ci\'c, Observation of trapped light within the radiation continuum, Nature {\bf 499}, 188 (2013).

\bibitem{Monticone2014}
F. Monticone and A. Al\'u, Embedded photonic eigenvalues in 3D nanostructures, Phys. Rev. Lett. {\bf 112}, 213903 (2014).

\bibitem{rybin2017optical}
 M. Rybin and Y. Kivshar, Supercavity lasing, Nature {\bf 541}, 164 (2017).

\bibitem{Balezin}
M. A. Belyakov, M. A. Balezin, Z. F. Sadrieva, P. V. Kapitanova, E. A. Nenasheva, A. F. Sadreev,  and A. A. Bogdanov, Experimental observation of symmetry protected bound state in the continuum in a chain of dielectric disks, arXiv preprint, arXiv:1806.01932 (2018).

\bibitem{Sadrieva2017}
Z.F. Sadrieva, I.S. Sinev, K.L. Koshelev, A. Samusev, I.V. Iorsh, O. Takayama, R. Malureanu, A.A. Bogdanov, and A.V. Lavrinenko, Transition from optical bound states in the continuum to leaky resonances: Role of substrate and roughness, ACS Photonics {\bf 4}, 723 (2017).

\bibitem{Plotnik2011}
Y. Plotnik, O. Peleg, F. Dreisow, M. Heinrich, S. Nolte, A. Szameit, and M. Segev, Experimental observation of optical bound states in the continuum, Phys. Rev. Lett. {\bf 107}, 183901 (2011).

\bibitem{molina2012surface}
M.I. Molina, A.E. Miroshnichenko, and Y.S. Kivshar, Surface bound states in the continuum, Phys. Rev. Lett. {\bf 108}, 070401 (2012).

\bibitem{corrielli2013observation}
G. Corrielli, G. Della Valle, A. Crespi, R. Osellame, and S. Longhi, Observation of surface states with algebraic localization, Phys. Rev. Lett. {\bf 111}, 220403 (2013).

\bibitem{rybin2017high}
M.V. Rybin, K.L. Koshelev, Z.F. Sadrieva, K.B. Samusev, A.A. Bogdanov, M.F. Limonov, Y.S. Kivshar, High-$Q$ supercavity modes in subwavelength dielectric resonators, Phys. Rev. Lett. {\bf 119}, 243901 (2017).

\bibitem{Fonda1963}
L. Fonda, Bound states embedded in the continuum and the formal theory of scattering, Ann. Phys. {\bf 22}, 123 (1963).

\bibitem{kim1999cs}
C.S. Kim, A.M. Satanin, Y.S. Joe, and R.M. Cosby, Resonant tunneling in a quantum waveguide: Effect of a finite-size attractive impurity, Phys. Rev. B {\bf 60}, 10962 (1999).

\bibitem{zhen2014topological}
B. Zhen, C.W. Hsu, L. Lu, A.D. Stone, and M. Solja\^ci\'c, Topological nature of optical bound states in the continuum, Phys. Rev. Lett. {\bf 113}, 257401 (2014).

\bibitem{fano}
M.F. Limonov, M.V. Rybin, A.N. Poddubny, and Y.S. Kivshar, Fano resonances in photonics, Nat. Photonics {\bf 11}, 543 (2017).

\bibitem{fano2016}
C. Blanchard, J. P. Hugonin, C. Sauvan. Fano resonances in photonic crystal slabs near optical bound states in the continuum. Phys. Rev. B {\bf 94}, 155303 (2016).

\bibitem{muljarov2017}
T. Weiss, M. Mesch, M. Schäferling, H. Giessen, W. Langbein, E. A.  Muljarov. From dark to bright: First-order perturbation theory with analytical mode normalization for plasmonic nanoantenna arrays applied to refractive index sensing. Phys. Rev. Lett.  {\bf 116}, 237401 (2016).

\bibitem{supp}
See Supplemental Material [link] for details on the rigorous derivation of the classical Fano formula for the transmission coefficient of a metasurface, the derivation of Eqs. (2)-(3), the results on eigenmode and reflection spectra dependence on the asymmetry parameter $\alpha$ for designs shown in Figs.1(b-f), the comparison of the analytical approach shown in Fig. 4 with numerical simulations using the realistic dispersion of silicon and the calculation of the BIC topological charge, which includes Refs.~\cite{sm1,sm2,sm3, sm4}.

\bibitem{sm1}
F. Alpeggiani, N. Parappurath, E. Verhagen, and L. Kuipers, Quasinormal-mode expansion of the scattering matrix, Phys. Rev. X {\bf 7}, 021035 (2017).

\bibitem{sm2}
A. B. Evlyukhin, C. Reinhardt, A. Seidel, B. S. Luk’yanchuk, and B. N. Chichkov, Optical response features of Si-nanoparticle arrays, Phys. Rev. B {\bf 82}, 045404 (2010).

\bibitem{sm3}
J. S. T. Gongora, G. Favraud, and A. Fratalocchi, Fundamental and high-order anapoles in all-dielectric metamaterials via Fano-Feshbach modes competition, Nanotechnology {\bf 28}, 104001 (2017).

\bibitem{sm4}
A. B. Evlyukhin, T. Fischer, C. Reinhardt, and B. N. Chichkov,  Optical theorem and multipole scattering of light by arbitrarily shaped nanoparticles. Phys. Rev. B {\bf 94}, 205434 (2016).

\bibitem{lee2012observation}
J. Lee, B. Zhen, S.-L. Chua, W. Qiu, J. D. Joannopoulos, M. Solja\^ci\'c, and O. Shapira, Observation and differentiation of unique high-$Q$ optical resonances near zero wave vector in macroscopic photonic crystal slabs, Phys. Rev. Lett. {\bf 109}, 067401 (2012).

\bibitem{thomas}
E Pshenay-Severin, A. Chipouline, J. Petschulat, U. H\"ubner, A. T\"unnermann, and T. Pertsch, Optical properties of metamaterials based on asymmetric double-wire structures, Opt. Express {\bf 19}, 6269 (2011).

\bibitem{shvets}
C. Wu, N. Arju, G. Kelp, J.A. Fan, J. Dominguez, E. Gonzales, E. Tutuc, I. Brener, and G. Shvets, Spectrally selective chiral silicon metasurfaces based on infrared Fano resonances, Nat. Commun. {\bf 5}, 3892 (2014).

\bibitem{costas2}
A. Jain, P. Moitra, T. Koschny, J. Valentine, and C.M. Soukoulis, Electric and magnetic response in dielectric dark states for low loss subwavelength optical meta atoms, Adv. Opt. Mater. {\bf 3}, 1431 (2015).

\bibitem{tian}
Y. Zhang, W. Liu, Z. Li, Z. Li, H. Chang, S. Chen, and J. Tian, High-quality-factor multiple Fano resonances for refractive index sensing, Opt. Lett. {\bf 43}, 1842 (2018).

\bibitem{shanhui}
V. Liu, M. Povinelli, S. Fan, Resonance-enhanced optical forces between coupled photonic crystal slabs, Opt. Express {\bf 17}, 21897 (2009).

\bibitem{costas}
A. Jain, P. Tassin, T. Koschny, and C.M. Soukoulis, Large quality factor in sheet metamaterials made from dark dielectric meta-atoms, Phys. Rev. Lett. {\bf 112}, 117403 (2014).

\bibitem{eit}
J. Hu, T. Lang, Z. Hong, C. Shen, and G. Shi, Comparison of Electromagnetically-induced transparency performance in metallic and all-dielectric metamaterials, J. Lightwave Technol. {\bf 36}, 2083 (2018).


\end{thebibliography}
\end{document}